\documentclass[10pt]{article}
\usepackage{epsfig}

\begin{document}

\title{\LARGE \bf Complete Classification of the String-like
Solutions \\ of the Gravitating Abelian Higgs Model\\}

\author{
\large M. Christensen$^a$\thanks{Electronic address: mc@bose.fys.ou.dk}$\:$,
A.L. Larsen$^a$\thanks{Electronic address: all@fysik.ou.dk} $\:$ and
Y. Verbin$^b$\thanks{Electronic address: verbin@oumail.openu.ac.il}}
\maketitle

\centerline{$^a$ \em Department of Physics, University of Odense, }
\centerline{\em Campusvej 55, 5230 Odense M, Denmark}
\vskip 0.4cm
\centerline{$^b$ \em Department of Natural Sciences, The Open University
of Israel,}
\centerline{\em P.O.B. 39328, Tel Aviv 61392, Israel}
\vskip 1.1cm

\begin{abstract}
The static cylindrically symmetric solutions of the
gravitating Abelian Higgs model form a two parameter family. In this paper
we give a complete
classification of the string-like solutions of this system. We show that
the parameter plane is
composed of two different regions with the following characteristics: One
region contains the
standard asymptotically conic cosmic string solutions together with a
second kind
of solutions with
Melvin-like asymptotic behavior. The other region contains two types of
solutions with bounded
radial extension. The border between the two regions is the curve of maximal
angular deficit of
$2\pi$. \\
\end{abstract}


\section{Introduction}
\setcounter{equation}{0}
\label{secintro}

Cosmic strings \cite{vilsh,KibbleH} were introduced into cosmology by
Kibble  \cite{Kibble1}, Zel'dovich \cite{Zel} and Vilenkin \cite{Vil1} as
(linear) topological  defects, which may have been formed during phase
transitions in the early  universe. Cosmic strings are considered as possible
sources for density perturbations and hence for structure formation in the
universe.

The simplest and most common field-theoretical model, which is used in order to
describe the generation of cosmic strings during a phase transition, is the
Abelian Higgs model. This model is known to have (magnetic) flux tube solutions
\cite{NO}, whose gravitational field is represented by an asymptotically conic
geometry. These are the local cosmic string solutions.

This conic space-time is a special case of the general  static and
cylindrically-symmetric vacuum solution of Einstein equations
\cite{exact-sol}, the so-called Kasner solution:
\begin{eqnarray}
ds^{2} = (kr)^{2a}dt^{2} - (kr)^{2c}dz^{2} - dr^{2}
- \beta^{2} (kr)^{2(b-1)}r^{2}d{\phi}^2
\label{Kasner1}
\end{eqnarray}
where $k$ sets the length scale while $\beta$ represents the asymptotic
structure, as will
be discussed below. Notice also that $(a, b, c)$ must satisfy the Kasner
conditions:
\begin{equation}
a + b + c=a^2 + b^2 + c^2=1
\label{Kasner2}
\end{equation}

The standard conic cosmic string solution \cite{Vil1,Garfinkle1} is
characterized
by an
asymptotic behavior given by (\ref{Kasner1}) with $a=c=0,\; b=1$ which is
evidently locally flat. In this case, the parameter $\beta$ represents a
conic angular
deficit \cite{Marder2,Bonnor}, which is also related to the mass distribution
of the source.

A first approximation to the relation between the
angular deficit $\delta\phi = 2 \pi (1-\beta)$ and the ``inertial mass" (per
unit length) $\tilde{m}$, of a local string was found
to be \cite{Vil1,Gott,Hiscock,Linet1}:
\begin{equation}
\delta\phi = 8 \pi G\tilde{m}
\label{angdef}
\end{equation}
Further corrections to (\ref{angdef}) were calculated in the following papers
\cite{LagMatz,GarfinkleLag,LagGarfinkle,CNV,Ver}.
\vskip 6pt
In order to fully analyse the string-like solutions of the Abelian Higgs
system  one
needs to solve the full system of coupled field equations for  the
gravitational
field and  matter fields (scalar + vector). But this is not necessary in
order to
obtain the asymptotic geometry. It is sufficient to note that for cylindrical
symmetry the flux tube has the property of ${\cal T}^0_0={\cal T}^{z}_{z}$,
where
${\cal T}^\mu_\nu$ is the energy-momentum tensor. This means that the
solution will
have a symmetry under boosts along the string  axis, i.e., $a=c$. The Kasner
conditions (\ref{Kasner2}) then  leave only two options; either the locally
flat
case:
\begin{equation}
a=c=0 \;\;\; , \;\;\; b=1
\label{SL}
\end{equation}
which we shall refer to as the cosmic string branch, or
\begin{equation}
a=c=2/3\;\;\; ,\;\;\;b=-1/3
\label{MW}
\end{equation}
which is the same behavior as that
of the Melvin solution \cite{Melvin}. We shall therefore refer to (\ref{MW}) as
the Melvin branch.

The Melvin branch does not provide
the same characteristics used to describe a ``standard" cosmic
string, and thus it has been
disregarded in previous investigations.
However, the Melvin-like solution seems completely well-behaved and one may
wonder whether it exists at all (asymptotically), or maybe it is just an
artifact of the too general reasoning above. Although several authors
\cite{CNV,Ver,Ortiz,FIU,Rayc} explicitly mention this possibility of a second
Melvin branch of solutions in the Abelian Higgs system, it has never been
properly discussed.

In this paper we show that indeed the Melvin branch actually exists
in the Abelian Higgs model. It turns out that each conic cosmic string
solution has a ``shadow" in the form of a corresponding solution in the Melvin
branch.
These two solutions, conic cosmic string and asymptotic Melvin, are found
only in
a part of the
two dimensional parameter space of the system, bounded by the curve of
maximal angular deficit of
$2\pi$.

There also seem to be some open questions in the literature
\cite{LagGarfinkle,Ortiz} about
the nature of the solutions beyond the border of maximal angular deficit of
$2\pi$. We will see that also in this region two types
of solutions coexist, both of them with a finite radial extension.\\

\section{General analysis of the Abelian Higgs system}
\setcounter{equation}{0}
\label{secgeneral}

The action of the gravitating Abelian Higgs model is:

\begin{equation}
S = \int d^4 x \sqrt{\mid g\mid } \left({1\over 2}D_{\mu}\Phi ^
{\ast}D^{\mu}\Phi
- {{\lambda  }\over 4}(\Phi ^{\ast} \Phi - v^2)^2 -
{1\over 4}{F}_{\mu \nu}{F}^{\mu \nu} + \frac{1}{16\pi G}
{\cal R}\right)
\label{higgsaction}
\end{equation}
where ${\cal R}$ is the Ricci scalar, ${F}_{\mu \nu}$ the Abelian field
strength, $\Phi$ is a complex scalar field with vacuum expectation value $v$
and $D_\mu = \nabla _{\mu} - ieA_{\mu}$ is the usual gauge covariant
derivative. We use units in which $\hbar=c=1$.

Because of the cylindrical symmetry of the source, we will use a line element
of the form:
\begin{equation}
ds^{2} = N^{2}(r)dt^{2} - d{r}^{2} - L^{2}(r)d{\phi}^2 - K^{2}(r)dz^{2}
\label{lineelement}
\end{equation}
and the usual Nielsen-Olesen ansatz for the +1 flux unit:
\begin{eqnarray}
\Phi=vf(r)e^{i\phi} \hspace{10 mm} , \hspace{10 mm}
A_\mu dx^\mu = {1\over e}(1-P(r))d\phi
\label{NOansatz}
\end{eqnarray}
This gives rise to the following field equations for the Abelian Higgs flux
tube:
\begin{eqnarray}
\frac{(NKLf')'}{NKL} + \left(\lambda v^2 (1-f^2) -
\frac{P^2}{L^2}\right)f = 0 \\
\frac{L}{NK}\left(\frac{NK}{L} P'\right)' - e^2 v^2 f^2 P = 0
\label{fluxtube}
\end{eqnarray}
With the line element (\ref{lineelement}), the components of the Ricci tensor
are:
\begin{eqnarray}
{\cal R}^{0}_{0} = - \frac{(LKN')'}{NLK} & , &
{\cal R}^{r}_{r} = - \frac{N''}{N} - \frac{L''}{L} - \frac{K''}{K}
\nonumber \\
{\cal R}^{\phi}_{\phi} = - \frac{(NKL')'}{NLK} &,&
{\cal R}^{z}_{z} = - \frac{(NLK')'}{NLK}
\label{Ricci}
\end{eqnarray}
The source is described by the energy-momentum tensor with the
following components:
\begin{eqnarray}
     {\cal T}^{0}_{0} &=& \rho = \varepsilon _s +\varepsilon _v +
     \varepsilon _{w} + u  \nonumber \\
     {\cal T}^{r}_{r} &=& -p_r = -\varepsilon _s -\varepsilon _v +
     \varepsilon _{w} + u \nonumber \\
     {\cal T}^{\phi}_{\phi} &=& -p_{\phi} = \varepsilon _s -
     \varepsilon _v -\varepsilon_{w} + u \nonumber \\
     {\cal T}^{z}_{z} &=& -p_z = \rho
\label{Tmunu}
\end{eqnarray}
where:
\begin{eqnarray}
\varepsilon_s = {v^2 \over 2} f'^2 \hspace{5 mm},\hspace{5 mm}
\varepsilon_v = \frac{P'^2}{2e^2 L^2} \hspace{5 mm},\hspace{5 mm}
\varepsilon_{w} = \frac{v^2 P^2 f^2}{2L^2} \hspace{5 mm},\hspace{5 mm}
u = \frac{\lambda v^4}{4} (1-f^2)^2
\label{densities}
\end{eqnarray}
It turns out to be convenient to use Einstein equations in the form (${\cal
T}\equiv {\cal
T}^{\lambda}_{\lambda}$):
\begin{equation}
{\cal R}_{\mu\nu} = -8 \pi G({\cal T}_{\mu \nu} -
\frac{1}{2} g_{\mu \nu}{\cal T})
\label{EinsteinR}
\end{equation}
 from which one obtains:
\begin{eqnarray}
\frac{(LKN')'}{NLK} &=& 4 \pi G(\rho +p_r+p_{\phi} +p_z)=
8 \pi G(\varepsilon_v - u)
\label{Einst0}
\end{eqnarray}
\begin{eqnarray}
\frac{(NKL')'}{NLK} &=& -4 \pi G(\rho -p_r+p_{\phi} -p_z)
= - 8 \pi G(\varepsilon_v + 2 \varepsilon_{w} + u)
\label{Einstphi}
\end{eqnarray}
\begin{eqnarray}
\frac{(LNK')'}{NLK} &=& -4 \pi G(\rho -p_r-p_{\phi} +p_z)=
8 \pi G(\varepsilon_v - u)
\label{Einst3}
\end{eqnarray}
and instead of the ``radial" part of (\ref{EinsteinR}), we take the following
combination:
\begin{eqnarray}
\frac{N'}{N} \frac{L'}{L} + \frac{L'}{L} \frac{K'}{K}+
\frac{K'}{K} \frac{N'}{N} = 8 \pi G p_r =
8 \pi G(\varepsilon _s +\varepsilon _v - \varepsilon _{w} - u)
\label{constraint}
\end{eqnarray}
which is not an independent equation but serves as a constraint. In vacuum, the
right-hand-sides of these equations vanish, and the first three of them are
trivially integrated. In this way we may get back Kasner's line element
(\ref{Kasner1}). This is therefore the asymptotic form of the metric tensor
around any (transversally) localized source and especially around an Abelian
Higgs flux tube.
Moreover, it is easy to get convinced that due to the symmetry under boosts
along the string axis, $K=N$.

The equations become more transparent if we express all lengths in terms of the
scalar characteristic length scale $1/\sqrt{\lambda v^2}$ (the ``correlation
length" in the superconductivity terminology). We therefore change to the
dimensionless length
coordinate $x=\sqrt{\lambda v^2}r$ and we introduce the metric component
$L(x)=\sqrt{\lambda
v^2}L(r)$. We also introduce the two parameters
$\alpha=e^2/\lambda$ and $\gamma=8\pi Gv^2$. In terms of these new
quantities, we
get a two parameter system of four coupled non-linear ordinary differential
equations (the prime now denotes $d/dx$):

\begin{eqnarray}
\frac{(N^2 Lf')'}{N^2 L} + \left(1-f^2 - \frac{P^2}{L^2}\right)f = 0
\label{systemNO1}\\
\frac{L}{N^2}\left(\frac{N^2 P'}{L}\right)' - \alpha f^2 P = 0
\label{systemNO2}
\end{eqnarray}
\begin{eqnarray}
\frac{(LNN')'}{N^2 L} &=&\gamma\left(\frac{P'^2}{2\alpha L^2} - \frac{1}{4}
(1-f^2)^2\right)
\label{systemE1}\\
\frac{(N^2 L')'}{N^2 L}
&=& - \gamma\left(\frac{P'^2}{2\alpha L^2} + \frac{P^2 f^2}{L^2} + \frac{1}{4}
(1-f^2)^2\right)
\label{systemE2}
\end{eqnarray}
We have also to keep in mind the existence of the constraint
(\ref{constraint}),
which gets the
following form:
\begin{eqnarray}
\frac{N'}{N} \left(2\frac{L'}{L} + \frac{N'}{N}\right) =
\gamma\left(\frac{f'^2}{2} + \frac{P'^2}{2\alpha L^2} -
\frac{P^2 f^2}{2L^2} - \frac{1}{4}(1-f^2)^2 \right)
\label{constraintmod}
\end{eqnarray}
In order to get string-like solutions, the scalar and gauge fields should
satisfy
the following boundary conditions:
\begin{eqnarray}
f(0)=0 &, & \lim_{x\rightarrow \infty} f(x) = 1 \nonumber \\
P(0)=1 &, & \lim_{x\rightarrow \infty} P(x) = 0
\label{boundarycond}
\end{eqnarray}
Moreover, regularity of the geometry on the symmetry axis $x=0$ will be
guaranteed by the
``initial conditions":
\begin{eqnarray}
L(0)=0 &, & L'(0) = 1 \nonumber \\
N(0) = 1 &, & N'(0) = 0
\label{initcond}
\end{eqnarray}

The purpose of the present paper is to map the two dimensional
$\alpha$-$\gamma$
parameter space, and thereby to classify all the string-like solutions of the
Abelian
Higgs system.
It is well-known that, even in flat space, the field equations can only be
solved numerically. However, much can be said about the nature of the solutions
even without explicitly solving the field equations
\cite{Garfinkle1,LagGarfinkle,CNV,FIU}.

The standard  Abelian Higgs
string solution, usually  considered in the literature, has a vanishing
gravitational mass. This
means that the spacetime around the string is locally flat except in the
core of the string, while
there is a non-trivial global effect namely a conical structure of the
space which is quantified by
an angular deficit. However, this does not at all saturate all the
possibilities of solutions of
(\ref{systemNO1})-(\ref{systemE2}). There are further types of solutions
with the same
boundary and ``initial" conditions, (\ref{boundarycond})-(\ref{initcond}),
which
are not asymptotically  flat but have interesting physical interpretations.
In this paper, we
will show that a point in the $\alpha$-$\gamma$ plane always represents two
solutions, except at the curve representing  angular deficit of $2 \pi$.
The various solutions are distinguished by, among other things,
their asymptotic geometries. \\

For analysing the solutions and obtaining the above-mentioned features and some
additional ones, we
introduce the Tolman mass (per unit length), $M$:
\begin{equation}
GM=2\pi G\int _{0}^\infty dr\; N^2 L(\rho +p_r+p_{\phi}
+p_z)=\frac{\gamma}{2} \int _{0}^\infty
dx\;  N^2 L \left(\frac{P'^2}{2\alpha L^2} - \frac{1}{4} (1-f^2)^2\right)
\label{mass}
\end{equation}
Using the field equations, one can show that \cite{Ver}:
\begin{equation}
GM=\frac{1}{2} \lim_{x\rightarrow\infty} (LNN')
\label{tolmanmass}
\end{equation}
We also define the magnetic field $B$:
\begin{equation}
B =  -\frac{1}{eL(r)}\frac{dP(r)}{dr}=-\frac{\gamma}{\alpha}\left(
\frac{e}{8\pi
G}\right)
\frac{P'(x)}{L(x)}
\label{magn}
\end{equation}
and the dimensionless parameter ${\cal B}=8\pi GB(0)/e$.
We find that the central value of the magnetic field (its value in the core
of the
string) can be expressed as \cite{Ver}:
\begin{equation}
{\cal B}= 1+2GM-\lim_{x\rightarrow\infty}(N^2 L')
\label{centralmag}
\end{equation}

The asymptotic form of the metric tensor is easily found by direct integration
of the two Einstein equations (\ref{systemE1})-(\ref{systemE2}) using the
boundary conditions and the definitions of $M$ and ${\cal B}$. It is  of
the Kasner
form
\begin{eqnarray}
N(x)=K(x)\sim \kappa x^a \hspace{6 mm},\hspace{6 mm} L(x)\sim \beta x^b
\label{asymptKasner1}
\end{eqnarray}
with:
\begin{eqnarray}
a=\frac{2GM}{6GM+1-{\cal B}} \hspace{6 mm},\hspace{6 mm}
b=\frac{2GM+1-{\cal B}}{6GM+1-{\cal B}} \hspace{6 mm},\hspace{6 mm}
\kappa ^2\beta=6GM+1-{\cal B}
\label{asymptKasner2}
\end{eqnarray}
The constant $\kappa$ appears free in the asymptotic solution, but it is
uniquely fixed in the complete one by the boundary conditions on the metric.

Due to (\ref{constraintmod}) we get the following relation
which is equivalent to the quadratic Kasner condition in Eq. (\ref{Kasner2}):
\begin{equation}
GM(3GM+1-{\cal B})=0
\label{MB}
\end{equation}
This immidiately points to the possibility of two branches of solutions
corresponding
to the vanishing of either factor in Eq. (\ref{MB}). These two possibilities of
course represent the two branches already discussed in Section 1.\\

Consider first the cosmic string branch ($a=c=0,\; b=1$) in a little more
detail. In this case, we find a simple physical interpretation to the
parameters
in Eq. (\ref{asymptKasner1}).
the constant $\beta=L'(\infty)$ defines the deficit angle
$\delta\phi=2\pi(1-\beta)$,
while the constant $\kappa=N(\infty)$ is the red/blue shift of
time between infinity and
the string core. Moreover, relations (\ref{tolmanmass}), (\ref{centralmag})
give:
\begin{equation}
M=0
\end{equation}
\begin{equation}
{\cal B}=1-\kappa ^2 \left(
1-\frac{\delta\phi}{2\pi}\right)
\end{equation}
That is to say, the Tolman mass vanishes and the central magnetic field is
directly expressed in
terms of the red/blue shift $\kappa $ and the deficit angle $\delta\phi$. \\

Then consider the Melvin branch ($a=c=2/3,\; b=-1/3$). In this case there
is no simple physical
interpretation of the parameters $\beta$ and $\kappa$ in Eq.
(\ref{asymptKasner1}).
As for
the Tolman mass, we
notice that it is non-zero, but the
equations (\ref{tolmanmass}), (\ref{centralmag}) lead to a simple
relation to the central
magnetic field
\cite{Ver}:
\begin{equation}
{\cal B}=1+3GM
\end{equation}
For more discussion of these general relations (and some other ones), we
refer to Ref. \cite{Ver}.
In the remaining sections of this paper, we shall construct explicitly the
various types of
solutions to (\ref{systemNO1})-(\ref{systemE2}) by using a relaxation
procedure
to solve the four
coupled differential equations numerically. The constraint
(\ref{constraintmod})
has been used for estimating the numerical errors. \\

\section{Open solutions: Cosmic strings and Melvin branch}
\setcounter{equation}{0}
\label{secopen}

\begin{figure}[hb]
\caption{Typical examples of the numerical solutions at low $\gamma$. Here, at
the point $(\alpha , \gamma )=(2,1.8)$ in parameter space, two solutions
coexist. The shown metric components $N$ and $L$  are for the cosmic string
branch and the Melvin branch respectively. Both solutions are open
and seem well behaved according to Eqs. (\ref{SL}) and (\ref{MW}).}
\centerline{\psfig{file=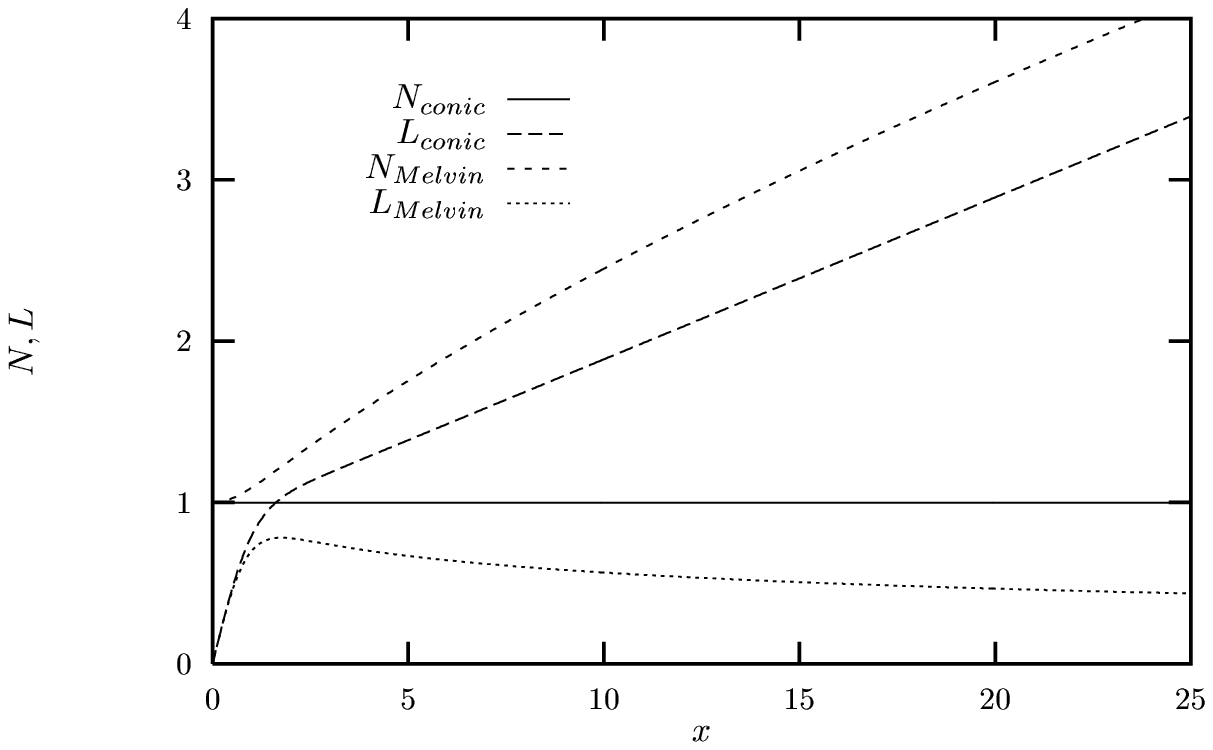,height=30cm,angle=0}}
\label{figopn}
\end{figure}

The existence of cosmic string solutions in the Abelian Higgs system is very
well established.
However, the existence of the solutions of the second type,
i.e. the Melvin branch type which is implied by (\ref{MW}) and (\ref{MB}), has
not been
properly
studied. We have found that for any conic cosmic string solution, an associated
solution exists in the Melvin branch. The main difference with respect to the
cosmic string branch is the fact that asymptotically the azimuthal circles have
vanishing circumference.
A related difference is the non-vanishing total mass (per unit
length) of the Melvin-like solutions. Figure \ref{figopn}
shows an example of the conic and the Melvin-like solution at the
point $(\alpha ,\gamma )=(2,1.8)$ in parameter space. Only the (square root of
the) metric components, $N$ and $L$, are plotted, as the scalar and vector
fields for the various solutions
deviate only very little from the standard and well examined cosmic
string configuration. The two branches are obviously quite different. The
cosmic string spacetime is asymptotically flat as $N$ is constant (actually
$N=1$ here as $\alpha = 2$ is the Bogomol'nyi limit \cite{Linet2,GibbCom})
and $L$
becomes
proportional to $x$ far from the core. In contrast hereto, the associated
solution in the Melvin branch has $N \propto x^{2/3}$ and $L \propto x^{- 1/3}$
in the asymptotic region. Therefore the circumference of circles lying in
planes
perpendicular to the core with $x=0$
as center will eventually decrease as one moves to larger $x$.

The curves shown in
Figure \ref{figopn} are representative for the cosmic string and Melvin
branches.
In the first
case, increasing
$\gamma$ at fixed $\alpha $ will
simply shift $N$ towards a higher constant value in the asymptotic region,
whereas $L$ will become less steep (meaning larger angular deficit). An
increase of $\alpha $ when $\gamma $ is held constant will have the opposite
effect, i.e. shifting $L$ to lower asymptotic values and decreasing the angular
deficit. This observation holds for all solutions (including the cylindrical
solutions and the closed solutions to be described in Section
\ref{secclosed}): only
the
$\gamma / \alpha$ ratio seems to matter for the asymptotic behavior. Still,
because of the region near the core, no obvious symmetry is observed that could
render the parameter space effectively one dimensional. In the case of the
Melvin branch, an increased $\gamma$ (again at fixed $\alpha$) tends to flatten
$N$, whereas the global maximum of $L$ takes a larger value at higher $x$. \\

Moving towards more massive strings in the parameter
space (larger $\gamma$ to $\alpha$ ratio) the angular deficit of the conic
solution will increase until it reaches a maximum of $2 \pi$.
The corresponding solution represents an asymptotically cylindrical manifold
with $N$ and $L$ both asymptotically constant.
One such solution exists for any $\alpha$ and the curve $\gamma _*(\alpha)$,
where $\delta \phi = 2 \pi$, plays a very special role in the
classification of
the various solutions. First of all, it obeys a power law \cite{LagGarfinkle}.
Secondly, as one
approaches a point on this curve in parameter space along the Melvin branch
the solution
converges to the same cylindrical solution as for the cosmic string; that
is, the conic solutions and the Melvin-like solutions coincide in a single
asymptotically cylindrical solution on the curve of maximal angular deficit.

\begin{figure}[hb]
\caption{The central magnetic field at the string core plotted for both the
cosmic
string and the Melvin branch. As $\gamma$ is increased (for fixed
$\alpha$), the
solutions in the two branches will converge towards the same cylindrical
solution
causing the two surfaces to intersect at the curve of angular deficit
$\delta \phi =
2 \pi$.}
\centerline{\psfig{file=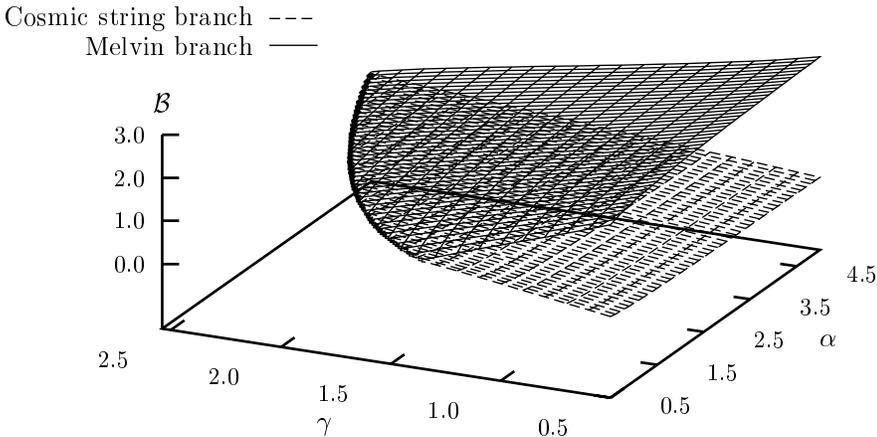,height=30cm,angle=0}}
\label{figmag}
\end{figure}

Figure \ref{figmag} illustrates how the central magnetic field
${\cal B}$ varies
throughout the parameter space for both the cosmic string and the Melvin-like
universes. Notice how the two surfaces intersect along the curve of maximal
angular deficit, where the two branches coincide in the cylindrical solutions.
\\

\section{Closed solutions: Inverted cone and singular Kasner solutions}
\setcounter{equation}{0}
\label{secclosed}

The issue of ``supermassive cosmic strings" in the Abelian Higgs system was
discussed at early days \cite{LagGarfinkle,Ortiz}. Two types of closed
solutions were
found. One \cite{LagGarfinkle} in which $N(x)$ vanishes at a finite distance
from the axis ($x=x_{max}$) while $L(x)$ diverges at that point. In the other,
$N(x)$ stays
finite for any $x$ while $L(x)$
decreases linearly outside the core of the string, and vanishes at a finite
distance from the axis. The geometry is still conic but of an inverted one
where the apex of the cone  is at the point where $L=0$ \cite{Gott,Ortiz}.
Using the same numerical methods as in the previous section, we have found
that
these two types of solutions are encountered just as the curve
$\delta\phi = 2 \pi$ is traversed. But the lack of asymptotic behavior for
closed solutions forces us to replace the boundary conditions
(\ref{boundarycond}) by the condition
\begin{eqnarray}
f(0)=0 &, & f(x_{max}) = 1 \nonumber \\
P(0)=1 &, & P(x_{max}) = 0
\label{bc2}
\end{eqnarray}
This ensures that the solutions are unit flux tubes, and makes the numerical
work simpler. For quite large radial extension this technicality is not of any
relevance as the variables $f$ and $P$ describing the scalar and vector fields,
respectively, converge quite fast to their values at ``infinity". \\

\begin{figure}[hb]
\caption{Typical examples of the super-massive solutions. Again two different
solutions coexist, here at $(\alpha , \gamma )=(2,2.05)$ which is above the
curve of maximal angular deficit. Both the inverted cone and the Kasner
solutions
represent closed solutions, and
for the Kasner-like solution $L$ has a singularity at $x_{max}$ where $N=0$. }
\centerline{\psfig{file=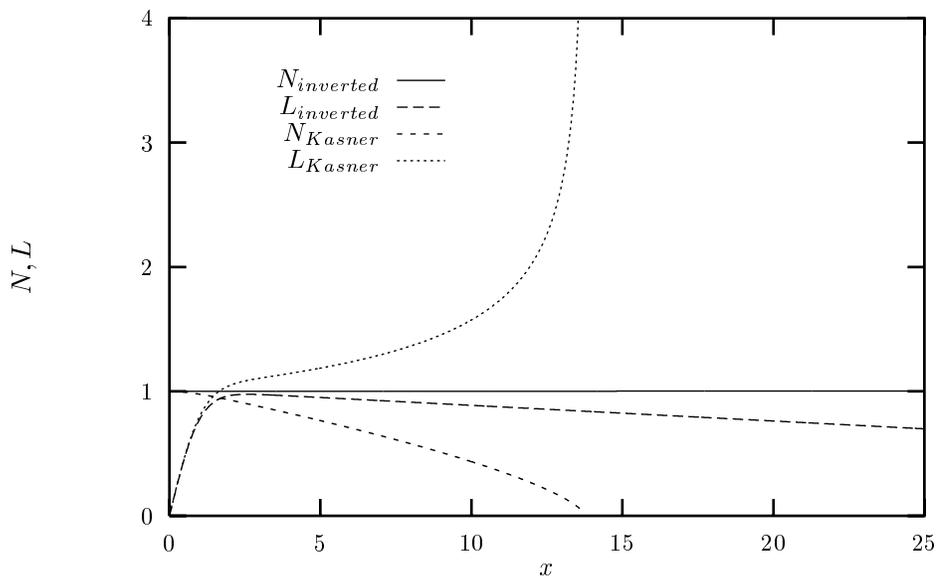,height=30cm,angle=0}}
\label{figclo}
\end{figure}

Figure \ref{figclo} shows an example of both
kinds of solutions at the point $(\alpha,\gamma)=(2,2.05)$ in parameter
space. This is
just above the curve of maximal angular deficit, which lies at $\gamma _*
(2)=2$ for fixed $\alpha =2$, and the solutions still have considerable sizes.
 Strictly speaking, these two types of solutions
have no asymptotic behavior as they are both representing closed solutions that
pinch off at a finite radial extension ($x_{max}$). But for slightly
supermassive configurations we have
\begin{eqnarray}
N(x)=K(x)\sim \kappa (x_{max} -x)^a \hspace{6 mm},\hspace{6 mm}
L(x)\sim \beta (x_{max} -x)^b
\label{singKasner1}
\end{eqnarray}
The inverted cone behaves according to Eq. (\ref{SL}) and
the singular Kasner solution behaves according to Eq. (\ref{MW}).

The curves shown in Figure \ref{figclo} are representative for the
solutions above
the curve of maximal angular deficit. As $\gamma$ increases ($\alpha $ fixed)
both solutions become smaller in radial extent. For the inverted cone $N$ gets
shifted to a lower constant value and $L$ intersects zero at smaller $x$.
Likewise $x_{max}$ decreases for the singular Kasner solution
causing $L$ to diverge at similarly low $x$. We have found that for any given
$\alpha $ and $\gamma > \gamma _* (\alpha)$, the solution with geometry as an
inverted cone has the larger size. \\

As mentioned, the solutions shrink when $\gamma $ is increased ($\alpha $
fixed). At some point the spacetime described by the solution becomes smaller
than a few times the characteristic core thickness which scales as $\alpha
^{-1/2}$ \cite{NO}. At even higher values of $\gamma$ the choice of
boundary conditions
becomes more and more important to fundamental questions as flux quantization
and topological stability; the solutions exist mathematically but are probably
unphysical.

\begin{figure}[hb]
\caption{Just above the curve of maximal angular deficit, which represents the
cylindrical solutions (lower curve), two closed solutions coexist. As $\gamma$
is increased (fixed $\alpha$) the radial extent of these solutions decreases.
The middle and upper curve represents respectively the singular Kasner-like
and
the inverted cone solutions with sizes of ten times the characteristic core
thickness. Above these two curves the solutions become unphysical. Notice that
the three curves can be fitted to power laws;
lower: $c_1 \approx 1.64 \; \wedge \; c_2 \approx 0.275 $,
middle: $c_1 \approx 1.72 \; \wedge \; c_2 \approx 0.285 $
and upper: $c_1 \approx 1.87 \; \wedge \; c_2 \approx 0.330 $.}
\centerline{\psfig{file=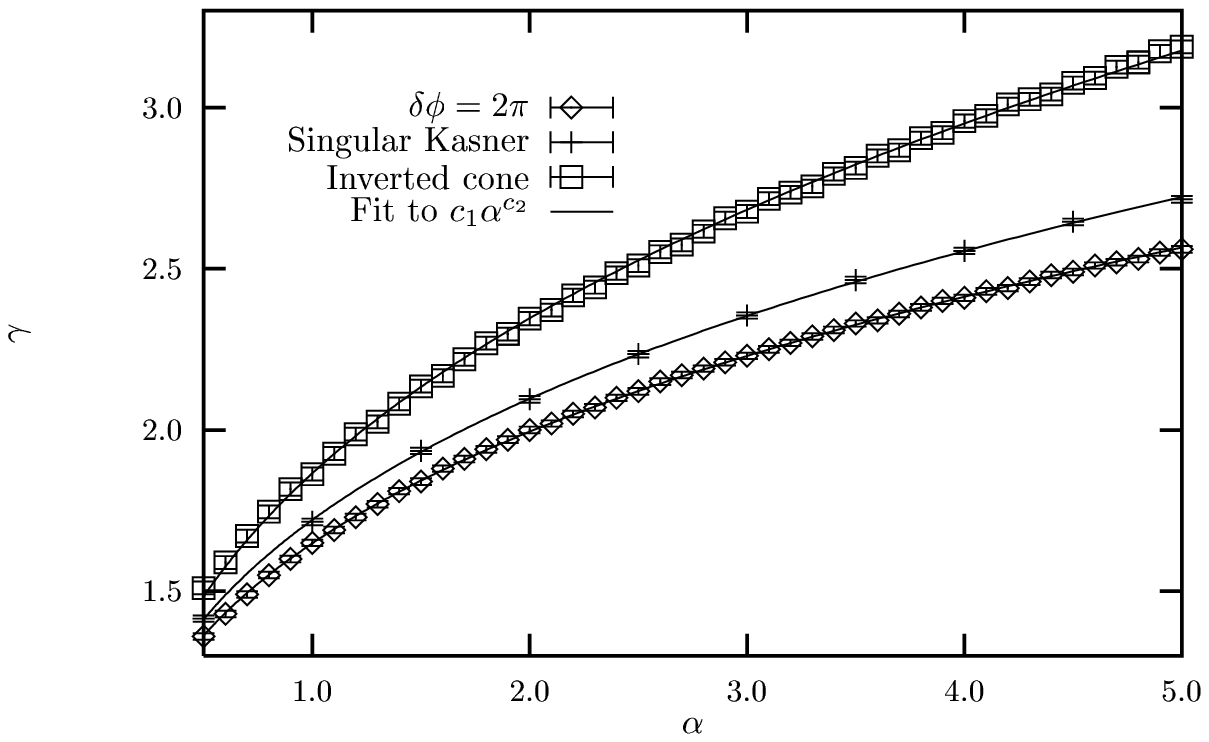,height=30cm,angle=0}}
\label{figlim}
\end{figure}

Figure \ref{figlim} shows three curves in parameter space which have all been
fitted to a power law, $\gamma = c_1 \alpha ^{c_2}$. The lowest one
is simply the curve of angular deficit $2 \pi$. This curve separates the two
open
solutions from the two closed ones, and represents itself a linear family of
asymptotically cylindrical solutions. The middle one represents the curve in
parameter space where the singular Kasner-like solutions have a radial extent
of $10 \, \alpha ^{-1/2}$, i.e. ten times the characteristic thickness of the
core. Similarly, the upper one represents the curve where the inverted cone
solutions pinches off at $x_{max} = 10 \, \alpha ^{-1/2}$. Starting on one of
these two curves and moving downwards (towards lower $\gamma$ for fixed
$\alpha$), the two closed solutions will become more and more similar and will
eventually coincide in the cylindrical solution. Moving
upward instead will yield rather small and increasingly unphysical
solutions. \\

\section{Conclusion}
\setcounter{equation}{0}

 Cylindrically symmetric configurations of the gravitating Abelian
 Higgs model have been
 examined by application of relaxation techniques to Einstein equations
 which simplify in this case to a system of coupled
 ordinary differential equations. Everywhere in parameter space two
 kinds of
 solutions exist, except at the curve of angular deficit $2\pi $, where
 only one
 cylindrical solution exists. The two open solutions are the standard
 asymptotically conic cosmic
 string and a solution with a Melvin like asymptotic behavior. The two
 closed solutions are the inverted
 cone and a singular Kasner-like solution. Supermassive solutions pinch
 off
 very fast and are probably unphysical. \\

 Here only the unit flux tube, $n=1$, has been considered, but
 introduction of an
 arbitrary $n \neq 0$ is expected to give similar families of
 solutions, which
 can be examined in an analogous way. \\

 Finally, it is worthwhile noting that the singularities of static
 field
 configurations generally seem to be lifted when time-dependence is
 reintroduced \cite{stringinfl}.  Therefore the nature of the singularities
of the supermassive strings
may be understood in the framework of time dependent analysis.


\noindent

\noindent

\noindent

\noindent


\begin{thebibliography}{123}

\bibitem{vilsh} A. Vilenkin and E.P.S. Shellard, {\it ``Cosmic strings and
other Topological Defects"} (Cambridge Univ. Press, Cambridge, 1994).
\bibitem{KibbleH} T.W.B Kibble and M. Hindmarsh,
Rep. Progr. Phys. {\bf 58}, 477 (1995).
\bibitem{Kibble1} T.W.B Kibble, J. Phys. {\bf A9}, 1387 (1976).
\bibitem{Zel} Ya. B. Zel'dovich, Mon. Not. R. Astron. Soc. {\bf 192}, 663
(1980).
\bibitem{Vil1} A. Vilenkin, Phys. Rev. {\bf D23}, 852 (1981).
\bibitem{NO} H.B. Nielsen and P. Olesen, Nucl. Phys. {\bf B61}, 45
(1973).
\bibitem{exact-sol} D. Kramer, H. Stephani, E. Herlt and M. MacCallum,
{\it ``Exact Solutions of Einstein's Field Equations"} (Cambridge Univ. Press,
 Cambridge, England 1980).
\bibitem{Garfinkle1} D. Garfinkle, Phys. Rev. {\bf D32}, 1323 (1985).
\bibitem{Marder2} L. Marder, Proc. Roy. Soc. London A {\bf252}, 45 (1959).
\bibitem{Bonnor} W.B. Bonnor, J. Phys. A {\bf12}, 847 (1979).
\bibitem{Gott} J.R. Gott, Astrophys. J. {\bf288}, 422 (1985).
\bibitem{Hiscock} W.A. Hiscock, Phys. Rev. {\bf D31}, 3288 (1985).
\bibitem{Linet1} B. Linet, Gen. Relativ. Gravit. {\bf 17}, 1109 (1985).
\bibitem{LagMatz} P. Laguna-Castillo and R.A. Matzner, Phys. Rev.{\bf D36},
3663 (1987).
\bibitem{GarfinkleLag} D. Garfinkle and P. Laguna, Phys. Rev. {\bf D39}, 1552
(1989).
\bibitem{LagGarfinkle} P. Laguna and D. Garfinkle, Phys. Rev. {\bf D40}, 1011
(1989).
\bibitem{CNV} J. Colding, N.K. Nielsen and Y. Verbin, Phys. Rev. {\bf D56},
3371 (1997).
\bibitem{Ver} Y. Verbin, {\it ``Cosmic Strings in the Abelian Higgs Model with
Conformal Coupling to Gravity"}, preprint hep-th/9809002, to appear in Phys.
Rev. {\bf D}.
\bibitem{Melvin} M.A. Melvin, Phys. Lett. {\bf 8}, 65 (1964).
\bibitem{Ortiz} M.E. Ortiz, Phys. Rev. {\bf D43}, 2521 (1991).
\bibitem{FIU} V.P. Frolov, W. Israel and W.G. Unruh, Phys Rev. {\bf D39},
1084 (1989).
\bibitem{Rayc} A.K. Raychaudhuri, Phys. Rev. {\bf D41}, 3041 (1990).
\bibitem{Linet2} B. Linet, Phys. Lett. A {\bf 124}, 240 (1987).
\bibitem{GibbCom} A. Comtet and G.W. Gibbons, Nucl. Phys. {\bf B299}, 719
(1988).
\bibitem{stringinfl} I. Cho, Phys. Rev. {\bf D58}, 103509 (1998).
\end{thebibliography}
\end{document}